\documentstyle [11pt]{article}
\newcommand{\beq}{\begin{equation}}   
\newcommand{\eeq}{\end{equation}}     %

\title{ Control Theoretic  Formulation of  Capacity of Dynamic Electro Magnetic
Channels }
\author{ {\bf  N.U.Ahmed, F. Rezaei and S. Loyka} \\ School of Information Technology and Engineering  \\
University of Ottawa, Ottawa\\
Ontario, Canada\\
\date{    } }
\begin{document}
\maketitle

\baselineskip 7.5 mm\noindent{\bf Abstract} In this paper  nonhomogeneous deterministic and stochastic Maxwell
equations  are used to rigorously formulate the capacity of electromagnetic   channels such as wave guides (
cavities, coaxial cables etc). Both distributed, but localized, and Dirichlet  boundary  data are considered  as
the potential input sources. We prove the existence of a source measure, satisfying certain second order
constraints (equivalent to power constraints), at which the channel capacity is attained. Further, necessary and
sufficient conditions for optimality are presented. \par\noindent {\bf Key words:} Maxwell Equations,
PDE, Semigroups, Stochastic Maxwell Equations, Channel Capacity.   \\
{\bf Subject Classification AMS(MOS):} 35Q60, 34H05, 94A40, 93Exx, 49K20
\section{Introduction}
Channel capacity for MIMO  channels (multiple input multiple output) has been the subject of intense study in
recent years. Most of the papers have been concerned with a strictly information theoretic analysis
\cite{telatar95, foschini98}. On the other hand, channel capacity can be treated as an optimization problem
subject to the constraint imposed by the Maxwell equations \cite{loyka2005}. In this paper, a mathematical
framework for MIMO capacity is provided using Electromagnetic constraints of the channel. It is shown  that this
problem can be rewritten as an optimal control problem where the control is the source measure subject to moment
constraints equivalent to transmitter power constraints. \par Rest of the paper is organized as follows:  The
current section ends after a brief list of notations. In section 2, we present the dynamic models of
electro-magnetic channels. Both distributed and boundary sources are considered, and existence and regularity
properties of solutions of the dynamic  systems   are presented.  In section 3, communication problems are
formulated. Section 4 deals with the solution of the problems proving existence of control measures from the
admissible class at which channel capacity is attained. In section 5, necessary and sufficient conditions of
optimality are  presented whereby a numerical  algorithm can be developed for capacity computation. The paper is
concluded with some comments in section 6. \vskip6pt \noindent {\bf Some notations:} Let $\Xi$ denote an
arbitrary set and ${\cal F}$ the Borel algebra of susbsets of the set $\Xi.$ We call the pair $(\Xi,{\cal F})$ a
measurable space. Let $\{\mu,\nu\}$ be any two regular Borel measures on the measurable space $(\Xi,{\cal F})$.
We let $\mu\prec \nu$ to denote the absolute continuity of the measure $\mu$ with respect to the measure $\nu.$
The Radon-Nikodym derivative, if it exists, of $\mu$ with respect to $\nu$ is denoted by $
\frac{\mu(dx)}{\nu(dx)} \equiv g(x),$ where $g \in L_1(\Xi,\nu).$
\par For any pair of Banach spaces $X,Y$, we let ${\cal L}(X,Y)$ denote the space  of bounded linear operators
from $X$ to $Y.$  For any bounded open connected domain $\Omega \subset R^n$ with sufficiently smooth boundary
$\partial \Omega,$ $H^s(\Omega,R^m) \subset L_2(\Omega,R^m), s\geq 0,$ will denote the standard Sobolev spaces
of functions defined on $\Omega$ and taking values from $R^m$ whose generalized derivatives up to order $s$
belong to $L_2(\Omega,R^m.$  Similarly $H^{-s}(\Omega,R^m), s \geq 0,$ will denote the Sobolev spaces with
negative exponents. These are distributions and, under some assumptions, are the topological duals of
$H^s(\Omega,R^m).$ By Sobolev's embedding theorem, it is known that for $s\geq (n/2) + k,$ $H^s(\Omega,R^m)
\hookrightarrow C^k(\Omega,R^m). $ Thus the Dirac measure $\delta_{\omega}(dx)$ with mass concentrated at
$\omega \in \Omega,$ $ b \delta_{\omega} \in H^{-s}(\Omega,R^m)$ for any  $b \in C(\Omega,R^m)$ and $s
>(n/2).$ Note that for $s \geq 0,$ a continuous linear functional  $\ell$ on $H^s(\Omega,R^m)$ has the
representation $$ \ell(\varphi) = \int_{\Omega} (\varphi,\psi) dx $$ for some $\psi \in H^{-s}(\Omega,R^m).$

For example,  for any $f \in L_2(\Omega,R^m)$  with $a_{\alpha}$ being constants, the function  $\psi,$ given by
$\psi \equiv \sum_{|\alpha|\leq s} a_{\alpha} D^{\alpha} f,$ is an element of $H^{-s}(\Omega,R^m).$ Here $\alpha
\equiv  (\alpha_1,\alpha_2,\cdots,\alpha_n)$ stands for the multi index and $|\alpha| = \sum_{i=1}^n
\alpha_i,\alpha_i \geq 0 $ and $D^{\alpha}f$ denotes the distributional derivative of $f$ of order $|\alpha|.$
For fractional $s,$ the Sobolev spaces are defined by use of Fourier transform.
\section{Channel Dynamics}
In this section we present  several models that describe the channel dynamics.  The first model is assumed to
satisfy  homogeneous Neumann  boundary condition (no leakage) with input source being  a vector of current  and
charge density.  The second model consists of  homogeneous wave equation  describing  the electric field with
nonhomogeneous  Dirichlet boundary  data whereby the input or source is provided.

 \subsection{\bf Channel with Current and charge as input Sources}  First we consider channels with homogeneous Neumann  boundary
condition. In this case the system is governed by a system of  wave equations arising from Lorenz transformation
of Maxwell's equations.  The  electromagnetic waves are  generated by input sources such as current  and charge
densities  and are confined in a wave guide. The electrical signals in the wave guide are governed by Maxwell's
equations. Using the vector and scalar potentials denoted by  $(a,\varphi)$ and the Lorentz gauge, the Maxwell's
equations  are  given by a system of wave equations:  \begin{eqnarray} &~& \partial^2 a/ \partial t^2  -  (1/\mu
\epsilon) \triangle a =(1/\epsilon )i, t \geq  0,\xi \in \Omega \subset R^3 \label{eqEM1}  \\ &~&
\partial^2\varphi/\partial t^2 - (1/\mu \epsilon) \triangle \varphi = (1/\mu \epsilon)   \rho ,\label{eqEM2} \end{eqnarray}
where $i$ and $ \rho$ are the sources, the first denoting the current density (vector) and the second the charge
density.  These are the sources that can be controlled to produce desirable field distributions inside the wave
guide. The field variables $\{E,B\}$  are related to  the potentials by the following equations:
\begin{eqnarray*} E = -(\dot a + \nabla \varphi), ~~ B =  \nabla \times a. \end{eqnarray*} These models are useful
in various fields of  communication such as radar,  optical fibre etc  \cite{ahmed03} (see references therein.).
We let $\Omega \subset R^3 $ denote an  open bounded  connected domain (representing the waveguide) having
piecewise smooth boundary.
\par Define  $ H \equiv L_2(\Omega,R^ 3)  \times L_2(\Omega,R) = L_2(\Omega,R^4)$ , denote $ y \equiv  (a,
\varphi)$ , and define the formal differential operator  $ C$  by  $ Cy \equiv (1/\mu \epsilon)(\triangle a,
\triangle \varphi) $ and let $B$ denote the Neumann  boundary operator and set $B(a,\varphi) =0 .$   This
operator is simply the outward normal derivative of the  arguments  at every point on the boundary of the wave
guide.   Then introduce the operator A as follows:
\begin{eqnarray*} &~& D(A) \equiv  \{  y  \in  H :  B(y) = 0 ~ \hbox{and}~ Cy \in  H\}   \subset   H ^2(\Omega
,R^3) \times H^2(\Omega ,R)
\\  &~&  \hbox{and set }  Az = Cz ~ \hbox{for}~  z \in  D(A).  \end{eqnarray*} Under  the given
boundary condition, $ -A $ is an unbounded positive self-adjoint operator in $H.$ Define the state space as  $
{\cal H}  \equiv  D(\sqrt -A) \times  H$  and the state as  $ z =(y, \dot y).$  This is the energy space.
Furnished with the scalar product and the associated  norm as presented below,\begin{eqnarray*} &~& (x,z)_{\cal
H} = (\sqrt -A x_1, \sqrt-A z_1)_{H} + (x_2,z_2)_{H} \\ &~&  \parallel x\parallel_{\cal H} \equiv \biggl(
\parallel \sqrt - A x_1\parallel_H^2 + \parallel x_2\parallel_H^2 \biggr)^{1/2},  \end{eqnarray*} ${\cal H}$ is
a Hilbert space. Note that the first term represents the potential energy and the second the kinetic energy
(magnetic field energy). Then we define the system  operator  ${\cal A}$  and the control operator ${\cal B}$ as
follows
\begin{eqnarray*} {\cal A} \equiv \biggl(\begin{array}{clcr}  0 & I \\ A & 0 \end{array} \biggr),
\end{eqnarray*}
\begin{eqnarray*} ~ {\cal B} \equiv  \left( \begin{array} {clcr} 0&0&0&0 \\ 0&0&0&0\\  0&0&0&0 \\ 0&0&0&0\\
(1/ \epsilon) &0&0&0 \\ 0 & (1/\epsilon) & 0 & 0 \\ 0&0&(1/ \epsilon) \\ 0&0&0& (1/\mu \epsilon)
\end{array} \right). \end{eqnarray*} Define the input or the control vector as $$ u \equiv
\left(\begin{array}{clcr} i_1 \\ i_2 \\  i_3 \\  \rho\end{array}\right). $$ Using these  notations the system of
wave equations given by (1) and (2) can be written as an abstract differential equation on the Hilbert space
${\cal H} $  as follows   \begin{eqnarray} \dot z = {\cal A}z + {\cal B}u, t \geq 0, \label {eqEM3}
\end{eqnarray}   where ${\cal  A }$  is an unbounded operator with domain and range in ${\cal H}.$  In practice
the input is localized. Let  $\Omega_0 \subset \Omega$ be a part of the domain at the input end of the wave
guide and consider the Hilbert space $U \equiv L_2(\Omega_0, R^4)$ with the standard scalar product. We may
assume that the  controls are functions of time taking values from the Hilbert space $U.$  Thus  our admissible
source is a proper subset ${\cal U}_{ad} \subset L_2(I,U).$  It can be shown that on the Hilbert space ${\cal
H}$ the system operator ${\cal A}$ is skew adjoint  and hence $i{\cal A}$ is self adjoint. Thus it follows from
semigroup theory (\cite{ahmed91},Theorem 3.1.4, p71),
in particular Stones theorem, that ${\cal A}$ generates a unitary group of operators ${\cal S}(t), t \in R . $
Using this unitary group of operators we can write the solution (mild) of equation (\ref{eqEM3}) as follows
\begin{eqnarray} z(t)= {\cal S}(t)z_0 + \int_0^t {\cal S}(t-s) {\cal B} u(s) ds, t \geq 0.\label{eqEM4} \end{eqnarray}
Note that in the absence of external input $u,$  the system is conservative and $$ \parallel
z(t)\parallel_{{\cal H}} = \parallel {\cal S}(t)z_0 \parallel_{{\cal H}}=\parallel z_0\parallel_{{\cal H}}
~\forall ~ t \in R.$$ This can be proved by simply  scalar multiplying in ${\cal H}$ on either side of the
equation
$$ \dot z = {\cal A}z , z(0) = z_0$$  by $z$ and integrating  and noting that
 $({\cal A}\xi,\xi)_{\cal H}= 0, \forall ~\xi \in D({\cal A}).$  We summarize  the above results in   the
 following theorem. \vskip6pt \noindent {\bf Theorem 2.1} For every input $u \in L_2(I,U)$ and initial state
 $z_0 \in {\cal H}$, the system (\ref{eqEM3}) has a unique mild solution $z \in C(I,{\cal H}).$ Further the solution
 is given by the expression (\ref{eqEM4}). This in turn implies that the system of wave equations
 (\ref{eqEM1})-(\ref{eqEM2}) has a unique mild  solution for every given initial state in the energy space and every
 given finite energy input.

\subsection{Channel with Dirichlet  Data  as Input Source} In cgs units, the Maxwell equations for electric field
$E$ and  magnetic field $B$ are given by\begin{eqnarray} &~& \nabla \times B = (1/c)
\partial E/\partial t + (4\pi/c) i \label {eqEM5} \\ &~& \nabla \times E = -(1/c) \partial B/\partial t,
\label{eqEM6}  \\ &~& \nabla \cdot E = 4\pi \rho,~ \nabla \cdot B = 0,  \label{eqEM7} \end{eqnarray}  where $c$
denotes the velocity of light and the pair $\{i,\rho\}$ denotes the current density vector and charge density
respectively. Here we have used standard notations for $ curl \phi \equiv \nabla \times \phi $ and $ div \phi
\equiv \nabla \cdot \phi .$ Using the first identity of equation (\ref{eqEM7}), the reader can easily verify
that $$ \nabla \times \nabla \times E = - \triangle E + 4\pi (\nabla \rho).$$ Now applying the curl operator on
either side of equation (\ref {eqEM6}) and using equation (\ref{eqEM5}) one can easily verify that the electric
field $E$ satisfies the following wave equation
\begin{eqnarray}  \partial^2 E /
\partial t^2 - c^2 \triangle E = - 4\pi (
\partial i/\partial t  + c^2 \nabla \rho ). \label {eqEM8} \end{eqnarray}  Since here we are interested in
boundary data, we assume that both the current and charge densities are identically zero. In order to solve such
equations in any bounded domain one must specify the initial and boundary  conditions. Hence the complete system
equation is given by \begin{eqnarray}  &~&  \partial^2 E / \partial t^2 - c^2 \triangle E = 0, \xi \in \Omega, t
\geq 0, \label {eqEM9}  \\  &~& E(0,\xi) = E_0(\xi) , \dot E(0,\xi) = E_1(\xi) , \xi \in \Omega, \label {eqEM10}
\\  &~& E(t,\xi)|_{\partial \Omega} = u(t,\xi), \xi \in \partial \Omega, t \geq 0. \label {eqEM11}
\end{eqnarray}  This is a initial boundary value problem  with nonhomogeneous Dirichlet boundary condition.   In
general the source  $u$ carries the information  to be transmitted  over the wave guide channel $\Omega.$ \par
There are two possible ways of attacking this problem. One is the semigroup approach (\cite{ahmed88}) and the
other is based on  the principle of transposition~(\cite{lions70}).
\par\noindent  {\bf Method A}
(Semigroup Approach): The first method is based on a well known technique (\cite{ahmed88},p59-63)
(and the references therein)  whereby one can transfer the boundary data to the righthand side of the original
differential equation. We write equation (\ref{eqEM9})-(\ref{eqEM11})  as a system
\begin{eqnarray} &~& \partial e/\partial t  = Le, \label{eqEM9A}
e(0) = e_0 \\ &~& Be = Tr  e_1 = u \label {eqEM9B}  \end{eqnarray} where $e \equiv (E,\dot E),$  $Tr \phi \equiv
\phi|_{\partial \Omega},$ and
\begin{eqnarray*} {L} \equiv \biggl(\begin{array}{clcr}  0 & I \\ c^2 \triangle    & 0 \end{array} \biggr). \end{eqnarray*}Note
that  this is a $6 \times 6$ matrix  with the elements of the first and the fourth diagonal blocks being all
zero and the second block being a $3\times3$ identity matrix and the third diagonal block is a $3\times 3$
diagonal matrix with the elements being  the Laplacian  $c^2 \triangle.$ To avoid introducing new notations, we
use the same symbols to define the operators $A, {\cal A}, B_0$ by
$$ A \equiv c^2  I \triangle|_{ker B}~~ \hbox{and}~~ {\cal A} \equiv L|_{ker B}~~\hbox{and}~~ B_0 \equiv B|_{ker
L}.$$  Note that the operator $A$  is a negative self adjoint unbounded operator  on  $ H \equiv
L_2(\Omega,R^3).$ The domain of the operator ${\cal A}$ is given by $$ D({\cal A}) = H^2(\Omega,R^3) \cap
H_0^1(\Omega,R^3) \times L_2(\Omega,R^3) \subset {\cal H} $$ where ${\cal H}$  is  the energy space, $$ {\cal H}
\equiv D(\sqrt -A)\times H,$$ considered here as the state space. Now returning to the system model, it is not
difficult to verify that the operator ${\cal A}$ is closed and
 densely defined and that    for any  $ R \ni \lambda \ne 0$,
 $$  \parallel R(\lambda,{\cal A}\parallel  \equiv  \parallel (\lambda I - {\cal
A})^{-1}\parallel_{{\cal H}}~ \leq ~(1/|\lambda|).$$ Thus by Hille-Yosida theorem (\cite{ahmed91},Theorem
2.2.8,p27)
, ${\cal A}$ is the infinitesimal generator of a $C_0$-group ${\cal S}(t), t \in R,$  of contractions in ${\cal
H}.$ Further, it is easy to verify that the operator ${\cal A}$ is skew adjoint and hence by Stones
theorem~(\cite{ahmed91},Theorem 3.1.4, p 71)
, it is the infinitesimal generator of a unitary group ${\cal S}(t), t \in R,$ on ${\cal H}.$ Our objective is
to convert the initial boundary value problem (\ref{eqEM9A})-(\ref{eqEM9B}) into a Cauchy problem (initial value
problem). Define
$${\cal W} \equiv H^2(\Omega,R^3)\times L_2(\Omega,R^3)\subset {\cal H}$$ and set ${\cal W}_1 \equiv Ker L ,
{\cal W}_2 \equiv Ker B.$ For any $\lambda( \in R)\ne 0$, define $ P \equiv R(\lambda,{\cal A}) (\lambda I -
L),$ and notice that $ P|_{{\cal W}_2} = I,$ the identity and that $P^2 = P.$ Thus ${\cal W}$ admits the direct
sum decomposition as follows, $$ {\cal W} = {\cal W}_1 \oplus {\cal W}_2.$$ Clearly $R(\lambda,{\cal A}) \in
{\cal L}({\cal H}, {\cal W}_2).$ For the source space, let $U$ be a linear subspace of $H^{3/2}(\partial
\Omega,R^3)$ carrying the structure of a Banach space  such that $B_0: {\cal W}_1 \longrightarrow U$ is
surjective and  ${\Re } \equiv (B_0)^{-1} \in {\cal L}(U,{\cal W}_1).$  Now going back to our original problem
(\ref{eqEM9A})-(\ref{eqEM9B}), we can rewrite the first equation in the equivalent form
\begin{eqnarray} \partial e/\partial t =  {\cal A}e  + (\Pi - (\lambda I - L))e  \label{eqEM9C}
\end{eqnarray} with $ \Pi \equiv (\lambda I -{\cal A}).$ For $\lambda \in \rho({\cal A})$, the resolvent set of
${\cal A}$, the operator $\Pi$ has bounded inverse giving the resolvent $R(\lambda,{\cal A}).$ Using the direct
sum decomposition, we can express the solution as  the sum given by $e = e^1+ e^2, e^1 \in {\cal W}_1, e^2 \in
{\cal W}_2.$ Substituting this in
 equation (\ref{eqEM9C}), and following similar  steps as presented in (\cite{ahmed88}, p59-62)
 , we arrive at the following abstract  Cauchy problem
\begin{eqnarray} &~&  \dot \zeta   = {\cal A} \zeta  + \Lambda {\Re} u, \zeta_0 \equiv  \zeta (0)
= R(\lambda,{\cal A})e_0,  \label{eqEM9D} \\
 &~& \Lambda  \equiv (I- R(\lambda,{\cal A})(\lambda I - L)), \label{eqEM9E} \\ &~&  e =
\Pi  \zeta . \label{eqEM9F} \end{eqnarray} Using the unitary group  introduced above, the  mild solution of the
system (\ref{eqEM9D})-(\ref{eqEM9E})- (\ref{eqEM9F})  is given by
\begin{eqnarray} &~& \zeta (t) =  {\cal S}(t) \zeta_0 +  \int_0^t {\cal S}(t-s) \Lambda {\Re}  u(s) ds, t \in I,
\label{eqEM9G}\\
&~& e(t) = \Pi \zeta (t), t \in I. \label{eqEM9H}\end{eqnarray}  Briefly  this  is the first method. It is clear
from the expression (\ref{eqEM9G}) that, for every $u \in L_2(I,U)$ and $ \zeta_0 \in D({\cal A}),$  $\zeta \in
C(I, D({\cal A}))$ and hence it follows from (\ref{eqEM9H}) that $e \in C(I,{\cal H}).$  We collect these facts
together in the following theorem.
\par \noindent {\bf Theorem 2.2} For every $e_0 \in {\cal H}$ and $u \in L_2(I,U)$,  the initial boundary value
problem (\ref{eqEM9A})-(\ref{eqEM9B}) has a unique mild solution  $e \in C(I,{\cal H}),$ and it is given by the
expressions (\ref{eqEM9G}) and (\ref{eqEM9H}).
\par\noindent  {\bf Method B} (Principle of Transposition):  The second method, which admits much more general boundary
data, is the method of transposition~(\cite{lions70}, p231, p283).
This method admits $L_2(\partial \Omega)$ data and, more generally, data from Sobolev spaces with negative norm
like $H^{-1/2}(\partial \Omega).$ The method consists of constructing a suitable isomorphism and then
transposing the isomorphism  for the solution of nonhomogeneous Dirichlet problems like
(\ref{eqEM9})-(\ref{eqEM11}). Consider the homogeneous Dirichlet problem
\begin{eqnarray}  &~& L\psi \equiv  \partial^2 \psi  / \partial t^2 - c^2 \triangle \psi = f,
\xi \in \Omega, t \geq 0, \label {eqEM15}  \\  &~& \psi(T,\xi) = 0 , \dot \psi(T,\xi) = 0, \xi \in \Omega,
\label {eqEM16}  \\  &~& \psi(t,\xi)|_{\partial \Omega} = 0, \xi \in \partial \Omega, t \geq 0, \label {eqEM17}
\end{eqnarray} for $f \in L_2(Q,R^3) \equiv L_2(I, L_2(\Omega,R^3)).$  Reversing the flow of time, it follows from the
results of the  previous subsection that for every $f \in L_2(Q, R^3)$ this problem has a unique solution $\psi
\in H^2(Q,R^3).$ Now  introduce the vector space $\Psi$ by \begin{eqnarray*} \Psi \equiv \biggl\{ \psi \in
L_2(Q,R^3): L\psi \in L_2(Q,R^3), \psi(T,\cdot) = 0, \dot \psi(T,\cdot) = 0, \psi|_{I\times \partial \Omega} = 0
\biggr\}\end{eqnarray*} and furnish it with the norm topology given by
\begin{eqnarray} \parallel \psi \parallel_{\Psi} \equiv \parallel L\psi \parallel_{L_2(Q,R^3)}\label{eqEM19} .\end{eqnarray}
The reader can easily verify that $\Psi$ is a  normed linear space. Since $L$ is a closed operator, it  follows
that $\Psi$ is a Banach space, in fact a Hilbert space. Thus it follows from the given norm topology that $L$ is
an iosmetric isomorphism of $\Psi$ onto $L_2(Q,R^3) \equiv L_2(I,L_2(\Omega,R^3)).$ For convenience of notation
we may express this fact by stating that  $$ L \in Iso(\Psi,L_2(Q,R^3)).$$ This  is known as the adjoint
isomorphism. Transposing this isomorphism, we can  settle the question of existence of solution of our original
nonhomogeneous boundary value problem (\ref{eqEM9})-(\ref{eqEM11}). This is stated in the following Theorem.
 \par\noindent {\bf Theorem 2.3} Consider the system (\ref{eqEM9})-(\ref{eqEM11}) and suppose $E_0 \in
H^{-1/2}(\Omega,R^3)$, $E_1 \in H^{-3/2}(\Omega,R^3)$ and $u \in L_2(I,H^{-1/2}(\partial \Omega,R^3)) \subset
H^{-1/2}(I\times \partial \Omega,R^3)$. Then the system (\ref{eqEM9})-(\ref{eqEM11}) has a unique solution $E
\in L_2(Q,R^3) = L_2(I,L_2(\Omega,R^3)).$ \vskip6pt\noindent {\bf Proof} Formally, scalar multiplying equation
(\ref{eqEM9}) by any $\psi \in \Psi $ and using Greens formula for integration by parts, one can easily derive
the following identity
\begin{eqnarray} &~& \int_{I\times \Omega} (E,L\psi) d\xi dt \nonumber  \\ &~& = \int_{\Omega}(E_1,\psi(0)) d\xi - \int_{\Omega}
(E_0,\dot \psi(0))d\xi - c^2 \int_{I\times \partial \Omega} (\partial \psi/\partial \nu,u(t,\xi) d\sigma(\xi)dt
,\label{eqEM19}\end{eqnarray} where $\partial \psi/\partial \nu$ denotes the partial derivative of $\psi$  in
the outward direction  of the unit normal vector $\nu$ at any position on the boundary $\partial \Omega$ and
$\sigma$ denotes  the surface (Lebesgue) measure on the boundary.  Define the functional
\begin{eqnarray} \ell(\psi) \equiv   \int_{\Omega}(E_1,\psi(0)) d\xi - \int_{\Omega}
(E_0,\dot \psi(0))d\xi - c^2 \int_{I\times \partial \Omega} (\partial \psi/\partial \nu,u(t,\xi)) d\sigma(\xi)dt
.\label{eqEM20}\end{eqnarray} Clearly this is a linear functional. Since $\psi \in H^2(Q,R^3)$, it follows from
standard trace theorems for Sobolev spaces  that  $\psi(0) \equiv \psi(0,\xi), \xi \in \Omega,$ is an element of
$H^{3/2}(\Omega,R^3),$  $\dot \psi(0) \in H^{1/2}(\Omega,R^3)$ and $\partial \psi/\partial \nu \in
H^{1/2}(I\times \partial \Omega,R^3).$ Thus for the given data $\{E_0,E_1,u\}$ with the regularities as
specified in the statement of the theorem, the scalar products on the righthand side of the identity
(\ref{eqEM20}) have the correct duality pairings. Hence we conclude that the given data determines a continuous
and hence a bounded linear functional  $\ell$ on the Banach space $\Psi.$ Since $ L \in Iso (\Psi, L_2(Q,R^3))$,
this means that  the composition map $(\ell o  L^{-1})$ is a continuous linear functional on $L_2(Q,R^3).$
Hence, by Riesz representation theorem, there exists an unique  $ E \in L_2(Q,R^3)$ such that  \begin{eqnarray}
(\ell o L^{-1})(f) = (E,f)_{L_2(Q,R^3)}~~\forall~~ f \in L_2(Q,R^3).\label{eqEM21} \end{eqnarray} Since $L$ is
an isomorphism, this is equivalent to saying that \begin{eqnarray} \ell(\psi) =
(E,L\psi)_{L_2(Q,R^3)}~~\forall~~ \psi \in \Psi .\label{eqEM21} \end{eqnarray} This also  verifies the validity
of the formal identity (\ref{eqEM19}) obtained by  integration by parts.  The uniqueness is a consequence of the
fact that $L \in Iso(\Psi,L_2(Q,R^3)).$  This completes the proof. $\bullet $
 \vskip6pt \noindent {\bf Remark 2.4} It
is clear from the above result that our nonhomogeneous Dirichlet initial boundary value problem
(\ref{eqEM9})-(\ref{eqEM11}) has a unique solution $E \in L_2(Q,R^3)$ for  a very general set of
 data form the class of generalized functions $$\{ E_0,E_1,u\} \in H^{-1/2}(\Omega,R^3) \times H^{-3/2}(\Omega,R^3) \times
 H^{-1/2}(I\times \partial \Omega,R^3).$$ For practical applications we  may limit our data from the Hilbert
 spaces $ L_2(\Omega,R^3)\times L_2(\Omega,R^3) \times L_2(I,L_2(\partial \Omega,R^3))$. In this case, off course, we expect
 our solutions to be much more regular or smooth.
\vskip6pt \noindent  Note that the data to solution map $\{E_0,E_1,u\} \longrightarrow E,$ which we denote by
$G,$ is a continuous linear map from  $H^{-1/2}(\Omega,R^3) \times H^{-3/2}(\Omega,R^3) \times H^{-1/2}(I\times
\partial \Omega,R^3)$ to $L_2(Q,R^3) \equiv L_2(I,L_2(\Omega,R^3))$ and hence there exists a
constant $K> 0$ such that \begin{eqnarray*} &~&  \hskip-30pt\parallel G(E_0,E_1,u)\parallel_{L_2(Q,R^3)} ~\leq \\
&~& K\biggl\{ \parallel E_0\parallel_{H^{-1/2}(\Omega,R^3)} + \parallel
E_1\parallel_{H^{-3/2}(\Omega,R^3)}\parallel + \parallel u\parallel_{H^{-1/2}( I\times \partial
\Omega,R^3)}\biggr\}. \end{eqnarray*} Since for $s>0$, the embeddings $ L_2(Q)\hookrightarrow H^{-s}(Q)$ are
continuous, it follows from the above result that for $\{E_0,E_1,u\} \in L_2(\Omega,R^3) \times
L_2(\Omega,R^3)\times L_2(I,L_2(\partial \Omega,R^3))$
\begin{eqnarray*} &~&  \hskip-30pt\parallel G(E_0,E_1,u)\parallel_{L_2(Q,R^3)} ~\leq \\ &~&
\tilde K \biggl\{ \parallel E_0\parallel_{L_2(\Omega,R^3)} + \parallel E_1\parallel_{L_2(\Omega,R^3)}\parallel +
\parallel u\parallel_{L_2( I, L_2(\partial \Omega,R^3)) }\biggr\}, \end{eqnarray*} where $\tilde K$ depends on
$K$ and the embedding constants $$ L_2(\Omega,R^3) \hookrightarrow
H^{-1/2}(\Omega,R^3),L_2(I,L_2(\partial\Omega,R^3)) \hookrightarrow  H^{-1/2}(I\times
\partial \Omega,R^3). $$

\vskip6pt\noindent In the study of communication problems we will set $E_0 = E_1 = 0$ and consider the boundary
data as the input source giving $E = G(u) \in L_2(I,H).$  A complete characterization of the input-output  map
is given in section 3. \vskip6pt \noindent {\bf Remark 2.5} It is interesting to note that Theorem 2.2 dealing
with the question of existence and regularity properties of solutions also provides a clue to numerical
technique for solving the basic problem (\ref{eqEM9})-(\ref{eqEM11}). Let $\{f_i\} \subset L_2(Q,R^3)$ be a
complete orthonormal set (orthonormality is not essential though linear independence is). Then note that
\begin{eqnarray*} (\ell o L^{-1})(f_i) = (E,f_i)_{L_2(Q,R^3)} \equiv c_i, i \in N. \end{eqnarray*}
These are precisely the
Fourier coefficients of $E$ with respect to the complete set $\{f_i\}$ as indicated by the righthand expression,
and hence $E$ is given by $ E = \sum_{i =1}^{\infty}  c_i f_i.$ Further it is clear that $c_i$,s are determined
 entirely by the data $\{E_0,E_1,u\} $ of the problem.
\vskip6pt\noindent  {\bf Remark 2.6} Comparing method A (Semigroup Approach)  with  method B (Principle of
Transposition), it is apparent that the later admits much more general data. At least for linear
initial-boundary value problems, semigroup theory seems to be  less powerful.

\section{ Formulation of Communication Problems}
\subsection{ Distributed Source:}{\bf   Transmit End  (T):} First we consider the system model described
by equation (\ref{eqEM3}) with the distributed source or control space $U = L_2(\Omega_0,R^4),$ that is, $$ U
\equiv \{ u \in L_2(\Omega,R^4): u(\xi) = 0, \forall ~ \xi \in \Omega \setminus \Omega_0\}.$$  For application
to communication problems we may simplify the source further by taking a finite number of disjoint closed
subsets $\{\sigma_i\}_{i=1}^n \subset \Omega_0$ and consider input source to be composed of the sum
\begin{eqnarray} u \equiv \sum_{i =1}^n x_i(t) \varphi_i(\xi), t \in I, \xi \in \Omega_0
\label{eqEM24}\end{eqnarray} where the functions $\varphi_i \in L_2(\Omega_0,R^4)$ vanishing outside $\sigma_i.$
 In other words, these functions have $\sigma_i$ as their supports and $x_i \in L_2(I)$ are scalar valued functions which are the
signals. These represent message signals radiated by the strategically located $n$-transmit antennas.
\vskip6pt\noindent {\bf Receiver End (R):} Let $S_0$ denote the receiving end of the wave guide. Sensors are
located on this set. Again let $\{\beta_i\}_{i=1}^m $ be a family of disjoint closed subsets of the set $S_0$
where the sensors are located. These sensors are assumed to be able to measure the electric field distribution
on these patches. These represent receiving antennas. In terms of the vector and scalar potential
$\{a,\varphi\}$ we have already seen that the electric field vector is given by
\begin{eqnarray*} E = -(\dot a + \nabla \varphi).
\end{eqnarray*} Hence in terms of the state variable we have
\begin{eqnarray} E \equiv -  \biggl(\begin{array}{clcr}  z_5+ \partial_1z_4  \\ z_6 + \partial_2z_4\\ z_7 + \partial_3 z_4
\end{array} \biggr) \equiv \Gamma z, \label{eqEM25} \end{eqnarray} where $\Gamma$ is the matrix of differential
operators easily determined by the above relation. Clearly, $\Gamma$ is a bounded linear operator from ${\cal
H}$ to $L_2(\Omega,R^3).$  The outputs are the integrals of weighted sensor response to the electric field
distribution on the patches. These are given by
\begin{eqnarray} y_i(t) = \int_0^t \biggl( \int_{\beta_i} <\alpha_i(\xi), E(s,\xi)>_{R^3} d\sigma(\xi)\biggr) ds
+ w_i(t),~ t \in I, i = 1,2\cdots,m ,\label{eqEM26}\end{eqnarray} where $\alpha_i$ is  the vector of  weight
given to the measured electric field distribution on  i-th site and  $w_i$ represents the measurement noise of
this site. The weight vector $\alpha_i$ may be assumed to be supported on the set $\beta_i.$   We assume that
$\{w_i\}_{i=1}^m$ are mutually independent standard Brownian motions with $W$ denoting the corresponding $m$
vector Brownian motion. Throughout the paper we use $(\Xi,{\cal F}, {\cal F}_t,P)$ to denote the filtered
probability space where ${\cal F}_t, t \geq 0,$ is an increasing family of right continuous subsigma algebras of
the sigma algebra ${\cal F}$. All random processes arising in this paper will be assumed to be based on this
complete filtered probability space.
\par  Now returning to our  problem and  using the source and the output models as described above, the state and the measurement
dynamics turn out to be  \begin{eqnarray} &~& \dot z = {\cal A} z  +  {\cal C}x, z(0) = z_0, \label{eqEM27}\\
&~& dy = {\cal G} \Gamma z dt + dW \label{eqEM28} \end{eqnarray} where the operators $\{ {\cal C}, {\cal G}\}$
are given by
$$ {\cal C} x \equiv  \sum_{i=1}^n x_i {\cal B}\varphi_i, ~\hbox{and}~ ({\cal G}_iE)(t) \equiv  \int_{\beta_i}
<\alpha_i(\xi),
 E(t,\xi)> d\sigma(\xi), i = 1,2,\cdots,m. $$ The reader can easily verify that the operators ${\cal C}: R^n \longrightarrow {\cal
 H}~\hbox{ and}~ ({\cal G} \Gamma): {\cal H} \longrightarrow R^m$  and they are bounded linear operators. Using the
 semigroup ${\cal S}(t),t\geq 0,$ corresponding to the operator ${\cal A},$ it follows from the expression
 (\ref{eqEM4}), that  \begin{eqnarray} z(t) = K_t(x) \equiv  {\cal S}(t)z_0 + \int_0^t {\cal S}(t-r){\cal C}x(r) dr
. \label{eqEMS1} \end{eqnarray} Define the composition map $F$ with values
  \begin{eqnarray} F_t(x) \equiv ({\cal G} \Gamma K_t)(x). \label{eqEMS2} \end{eqnarray}
 Clearly $F$ is a nonanticipative (causal) operator mapping $L_2(I,R^n)$ to $L_2(I,R^m)\cap C(I,R^m) $ and,
  being the composition of
 bounded linear operators, it is also a  bounded linear operator.   Thus the output
 equations
  (\ref{eqEM26}) can be written as as a linear  stochastic differential equation in $R^m$,
    \begin{eqnarray} dy = F_t(x) dt + dW, y(0) = 0,  t \in I. \label{eqEMS3}
  \end{eqnarray} In case of method B, the map $F$ is given by the composition map $F_t(x) = ({\cal G} G {\cal
  C})_ t(x),$ which is a bounded linear operator from $L_2(I,R^n)$ to $L_2(I,R^m).$

\subsection{Boundary Source:}  Next we consider the model described by the boundary value problem
(\ref{eqEM9})-(\ref{eqEM11}). Here we consider a part  $\partial \Omega_0 \subset \partial \Omega$ of the
boundary $\partial \Omega$ where the source is active.  Then for   the source space  we take  $L_2(I,U) $ where
$U$ is a closed linear  subspace of $ H^{3/2}(\partial \Omega_0,R^3) \equiv \{\varphi \in H^{3/2}(\partial
\Omega,R^3): \varphi (\xi) = 0 ~\hbox{for}~\xi \not \in \partial \Omega_0\}.$ Again we let $\{\sigma_i\} $
denote a family of disjoint subsets of the set $\partial \Omega_0$ and model the input as
\begin{eqnarray} u(t,\xi) \equiv \sum_{i=1}^n x_i(t) \psi_i(\xi), t \in I, \xi \in \partial
\Omega_0  \label{eqEM29}\end{eqnarray} where $\psi_i  \in  U $ and supported on the set $\sigma_i$ and $x_i \in
L_2(I).$ The complete system model is then given by
\begin{eqnarray}  &~&  \partial^2 E / \partial t^2 - c^2 \triangle E = 0,
\xi \in \Omega, t \geq 0, \label {eqEM30}  \\  &~& E(0,\xi) = 0 , \dot E(0,\xi) = 0 , \xi \in \Omega, \label
{eqEM31}  \\  &~& E(t,\xi)|_{\partial \Omega} = u(t,\xi) = \sum_{i=1}^n x_i(t) \psi_i(\xi), \xi \in \partial
\Omega, t \geq 0. \label {eqEM32} \end{eqnarray} For this model, we can use the representations
(\ref{eqEM9G})-(\ref{eqEM9H}) and (\ref{eqEM32}) to construct the output equation. Define the map
\begin{eqnarray} K_t(x) \equiv \Pi \biggl( {\cal S}(t)\zeta_0 + \int_0^t {\cal S}(t-r) \Lambda {\Re}{\cal C} x(r)
dr \biggr), t \in I. \label{eqEMS4} \end{eqnarray} Let $\Gamma$ denote the projection  map $\Gamma e \equiv e_1$
which projects $e \equiv  (e_1,e_2) =  (E,\dot E)$ to the first component $e_1 = E.$  Using these maps, again we
can write the output
equation  in the same general form  (\ref{eqEMS3}), \begin{eqnarray} dy  &=& ({\cal G}\Gamma K_t)(x) dt + dW \nonumber \\
&=& F_t(x) dt + dW, t \in I. \label{eqEMS5} \end{eqnarray} The existence of the map $F$ is  assured by the
expressions (\ref{eqEM9G})-(\ref{eqEM9H}) as presented in the semigroup approach (method A).
\subsection{Noisy Source}  So far we have assumed that the source is noise free. In order to admit noisy source one must
add some compatible  additional terms to the evolution equations (\ref{eqEM3}) and (\ref{eqEM9D}).  For the
distributed source, we replace the evolution equation  (\ref{eqEM3}) by the stochastic differential equation
\begin{eqnarray} dz = {\cal A}z dt + {\cal B}u dt + \sigma dW^o, t \geq 0, \label{eqES1} \end{eqnarray}
on the Hilbert space ${\cal H},$  where the operator $\sigma$ is given by a $8\times 4 $ matrix of operators
with the first four rows being all zero and the remaining $4\times 4$ matrix is a diagonal matrix of operators
$\{\sigma_i, i =1,2,3,4\}.$ The  Brownian motion $W^o$ is given by the vector $ W^o \equiv \hbox{col}
\{W_1^o,W_2^o,W_3^o,W_4^o\}$ of independent Brownian motions $$ W^o_i(t,\cdot) \equiv \{W^o_i(t,\xi),\xi \in
\Omega\} , i =1,2,3,4,$$ each  taking values possibly from $L_2(\Omega,R).$  In reference to the field equations
(\ref{eqEM1}) and (\ref{eqEM2}), this means adding distributed white noise on the righthand side of each of the
equations in the form $ \sigma_i \dot W^o_i(t)  \equiv \sigma_i(\cdot ) \dot W^o_i(t,\cdot), i = 1,2,3,4.$ This
model allows one to deal with  localized as well as distributed noise around the wave guide. Letting $V$ denote
any  separable Hilbert space, for example a closed linear subspace of  $L_2(\Omega,R^4),$  we may assume $W^o$
to be a
   $V$ valued Brownian motion,
independent of the Brownian motion $W$ (receiver noise), with covariance operator denoted by $Q^o$ and $\sigma
\in {\cal L}(V,{\cal H})$ so that $ Q^o_{\sigma} \equiv \sigma Q^o \sigma^*$ is a positive nuclear operator in
${\cal H}$.  A natural choice for the space $V$ is $L_2(\Omega_0,R^4) \equiv U,$ same as the source space,  and
$\sigma = {\cal B}.$ In any case this choice is determined primarily by physical requirements and mathematical
simplicities.  Since ${\cal S}(t),t \in R,$ is a unitary group, it is easy to verify that ${\cal
S}(t)Q^o_{\sigma}S^*(t)$ is a positive nuclear operator in ${\cal H}$ for all $t\in R$ given that $Q^o_{\sigma}$
is. Thus for each $u \in L_2(I,U),$ equation (\ref{eqES1}) has a unique mild solution $z,$ which belongs to  $
C(I,{\cal H})$  with probability one, possessing  bounded second moments. In this case the map $K_t(x), t \geq
0,$ is given by
\begin{eqnarray} K_t(x) \equiv  z(t) = {\cal S}(t)z_0 + \int_0^t {\cal S}(t-r){\cal C}x(r) dr + \int_0^t {\cal
S}(t-r) \sigma dW^o. \label{eqES2} \end{eqnarray}   \par Considering the boundary value problems,  for noisy
boundary source, equation (\ref{eqEM9D}) is replaced by \begin{eqnarray}  d\zeta  = ({\cal A}\zeta  + \Lambda
\Re u) dt + \Lambda \Re \sigma dW^o, t \geq 0. \label{eqES3} \end{eqnarray}   For  the boundary source we had
chosen $U \subset  H^{3/2}(\partial \Omega_0,R^3).$ Thus it is necessary that  $\sigma \in {\cal L}(V,U)$ where
$V$ is any separable Hilbert space supporting the Brownian motion $W^o.$ For example, $V \equiv L_2(\partial
\Omega_0,R^3)$ or any closed linear subspace thereof.  Again for the existence of mild solutions $e \equiv \Pi
\zeta \in C(I, {\cal H}),$ it suffices if $\Lambda \Re \sigma Q^o\sigma^* \Re^* \Lambda^*$ is a positive nuclear
operator in ${\cal H}.$ In this case, equation (\ref{eqEMS4}) is replaced; and the process $K_t(x), t \geq 0,$
is given by
\begin{eqnarray} K_t(x) \equiv \Pi \biggl( {\cal S}(t)\zeta_0 + \int_0^t {\cal S}(t-r) \Lambda {\Re}{\cal C}
x(r) dr  + \int_0^t {\cal S}(t-r) \Lambda {\Re} \sigma dW^o\biggr), t \in I. \label{eqES4} \end{eqnarray}
\vskip6pt \noindent {\bf Remark 3.1.} In case the sensors (receiving antennas) are nonlinear, the operator
${\cal G}$ is nonlinear and hence the composition map $F_t, t \in I,$ is also nonlinear. The results presented
in this paper remain valid provided this nonlinearity is  uniformly Lipschitz having at most linear growth.
\vskip6pt
\section{Channel Capacity}  In view of the preceding  discussions, we notice that the input and output spaces are
given by $X \equiv L_2(I,R^n)$ and  $Y \equiv C(I,R^m).$ Suppose these spaces are furnished with the
(topological) Borel algebra turning them into measurable spaces $(X,B_X)$ and $(Y,B_Y).$ Let $M(X)$  and $M(Y)$
denote the space of Borel probability measures on $(X,B_X)$ and $(Y,B_Y)$ respectively.  Considering the source
space, let $M_2(X) \subset M(X)$ denote the class of probability measures having  finite second moments, that
is, $$ \mu \in M_2(X) ~\hbox{if and only if}~ \int_{X} |x|_X^2~ \mu(dx) < \infty.$$ Since normally the source
power is limited, we consider a  bounded subset of $M_2(X)$ given by $$ S_r \equiv \{ \mu \in M_2(X): \int_X
|x|_X^2 ~ \mu(dx) \leq r T\}$$ where  $r>0$ is the power constraint and $T$ is   the length of the time interval
$I$ denoting the duration of the message source. For admissible source measures, we can choose  any set ${\cal
M}_{ad}$ which is a weakly compact and convex  subset of the set  $S_r$. For example, ${\cal M}_{ad} = wc\ell
(\Gamma_r)$ where $\Gamma_r$ is any  convex subset of the set  $S_r$ satisfying
$$ \lim_{n\rightarrow \infty} \sup_{\mu \in \Gamma_r} \int_X \sum_{i\geq n}(x,e_i)^2 \mu(dx) = 0$$ for any
orthonormal basis  $\{e_i\}$ of the Hilbert space $X.$ Under this  assumption, the set $\Gamma_r$ is uniformly
tight and hence conditionally weakly compact. Thus its weak closure is weakly compact. For more concrete
examples of compact sets ${\cal M}_{ad}$ see Remark 4.2 following theorem 4.1.
\par Considering the output space $(Y,B_Y),$ let $M(Y)$ denote the space of regular Borel probability measures
on it. Let $M(X\times Y)$ denote the space of joint Borel probability measures on the product sigma algebra $B_X
\times B_Y.$ We have seen in section 3, that the output signal $y$ is related to the input process $x$  through
the communication system (\ref{eqEM27})-(\ref{eqEM28}) leading to (\ref{eqEMS3}) for the distributed source; and
(\ref{eqEM30})-(\ref{eqEM32}) leading to (\ref{eqEMS5})  for the Dirichlet source.  In other words, for a given
probability measure $\mu \in S_r \subset M_2(X)$ on the input space, there is a unique measure $\nu \in M(Y)$ on
the output space $Y$ induced by the channel.  Let $\gamma \in M(X\times Y)$ denote the joint probability measure
and $\mu \times \nu$ the product measure with $\mu$ and $\nu$ being the marginals of $\gamma.$  The relative
entropy of $\gamma$ with respect to the product measure $\mu\times\nu,$ denoted by $I({\cal X},{\cal Y}),$  is
called the mutual information which is a measure of the amount of information carried by the observable  noisy
output ${\cal Y}$  about the
 input message (source) ${\cal X}$.  This is given by the following expression,
  \begin{eqnarray} I({\cal X},{\cal Y} ) \equiv \int_{X\times Y} log \biggl( \frac {\gamma(dx\times
dy)}{\mu(dx)\times \nu(dy)} \biggr)~ \gamma(dx\times dy), \label{eqEM34}\end{eqnarray} where  $ \Upsilon(x,y)
\equiv \frac {\gamma(dx\times dy)}{\mu(dx)\times \nu(dy)}$ denotes the Radon-Nikodym derivative of $\gamma$ with
respect to the product measure.  Clearly this requires that $\gamma$ be absolutely continuous with respect to
the product probability measure $\mu \times \nu.$ Note that the output measure $\nu$ is related to the input
measure $\mu$ through the channel operator and it is given by
\begin{eqnarray} \nu(D) = \gamma(X\times D) = \int_X q(x,D) \mu(dx), ~\forall ~ D\in B_Y \label{eqEM35}
\end{eqnarray} where $$ q(x,D) = Pr\{ y \in D| x\} $$ is the conditional probability of the output $y$ being in $D
\in B_Y$ given that the input  $x \in X .$  This is precisely the action of the channel on the input and, as we
have seen in the preceding sections, it is determined by the dynamic models of the channel.  A closed form
expression for this will follow shortly. Substituting the expression (\ref{eqEM35}) into the expression
(\ref{eqEM34}) we obtain
\begin{eqnarray} I({\cal X},{\cal Y}) \equiv J(\mu) \equiv \int_{X\times Y} log \biggl( \frac {q(x,dy)}
{\int_X q(\xi,dy)\mu(d\xi)}\biggr) q(x,dy) \mu(dx) \label{eqEM36}\end{eqnarray} which is a functional of the
 measures $q$ and $\mu.$   Since the channel dynamics is given, this is a functional of the
source measure only  as indicated above.  We have seen in the preceding section that, for both  the distributed
and the boundary sources, the output equation  has the general form given by  a linear  stochastic differential
equation in $R^m$,
\begin{eqnarray} dy = F_t(x) dt + dW, y(0) = 0, \label{eqEM37}\end{eqnarray} where $F$ is the causal (nonanticipative)  map  $F: L_2(I,R^n)
\longrightarrow L_2(I,R^m)$ as described earlier. This is a continuous linear map.  Now it  follows from
equation (\ref{eqEM37}) that for every given $x \in X$, $q(x,\cdot) $ is a Gaussian measure on $Y$ with mean
trajectory given by
\begin{eqnarray} \bar F(x) \equiv \{ \int_0^t F_s(x) ds, t \in I\} \in Y \label{eqEM38}
\end{eqnarray} while the covariance operator $Q_1$ is given by \begin{eqnarray*}  (Q_1\xi,\xi) \equiv
\int_{I^2}(K_1(t,s)\xi(s),\xi(t)) ds dt, \xi \in Y ,
\end{eqnarray*} with the kernel of the operator $Q_1$ being   $ K_1(t,s) \equiv (t\wedge s) I_m, (t,s) \in I\times I,$
and $I_m$ is the identity matrix of dimension $m.$    Thus the Channel Kernel is given by the conditional
Gaussian measure
\begin{eqnarray} q(x,D) \equiv N_G(\bar F(x),Q_1)(D), x \in X, D \in B_Y.  \label{eqEM39} \end{eqnarray}
Since $\bar F$ is a continuous linear map from $X$ to $Y$, and   $Range (\bar F(x)) \subset Range (Q_1^{1/2})$
for all $ x \in X,$ it is clear that,  for every $D \in B_Y$,  the map $x \longrightarrow q(x,D)$ is continuous
from $X$ to the interval $[0,1].$  Using the expression (\ref{eqEM39}) in the expression for the mutual
information given by (\ref{eqEM36}) we obtain the following equivalent expression,
\begin{eqnarray}  J(\mu) \equiv \int_{X\times Y} log \biggl( \frac {N_G(\bar F(x),Q_1)(dy)}
{\int_X N_G(\bar F(\xi),Q_1)(dy) \mu(d\xi)}\biggr) N_G(\bar F(x),Q_1)(dy) \mu(dx), \label{eqEM40}\end{eqnarray}
which is clearly dependent on the channel  operator $F.$ Denoting the convolution $$ \int_X N_G(\bar
F(x),Q_1)(D) \mu(dx) \equiv \nu_G(D), D \in B_Y$$ the reader can easily verify that $$ N_G(\bar F(x),Q_1)(\cdot)
\prec \nu_G(\cdot)$$ for $\mu$ almost all $x \in X.$ Thus the Radon-Nikodym derivative  of $N_G$ with respect to
the measure $\nu_G$ exists and hence $J(\mu)$ given by (\ref{eqEM40}) is well defined.  Now our objective is to
determine the capacity of the channel by maximizing the above functional  over a set of admissible measures on
the source space subject to  power constraints. That is we must find
\begin{eqnarray} C \equiv \sup \{ J(\mu), \mu \in  {\cal M}_{ad}\}  \label{eqEM41} \end{eqnarray}
 where the set ${\cal M}_{ad},$ as
defined before, is any weakly compact convex subset of the set
\begin{eqnarray}  S_r \equiv \{ \mu \in M_2(X): \int_X |x|_X^2 ~ \mu(dx) \leq r T\}. \label{eqEM42}
\end{eqnarray}   The first question that we must  address is: does the supremum exist
and, if  it does, is it attained on the set ${\cal M}_{ad}.$  Without much additional assumptions we can prove
the following result.
 \vskip6pt \noindent {\bf Theorem 4.1}  Suppose ${\cal M}_{ad}$  is a  weakly compact and convex subset of the set
  $S_r \subset M_2(X)$. Then, there  exists a unique  $\mu^o \in {\cal M}_{ad}$ at which $J$
attains its supremum. In other words capacity is attained. \vskip6pt \noindent {\bf Proof.} For the existence of
supremum on ${\cal M}_{ad}$ it suffices to prove that $J$ is weakly upper semicontinuos and bounded away from
$+\infty.$  For uniqueness we show that $J$ is strictly concave.  First we prove that the functional $\mu
\longrightarrow J(\mu)$ is concave and weakly upper semicontinuous on $M(X).$ For simplicity of notation we
revert back to our original notation and set $ N_G(\bar F(x),Q_1)(D) \equiv q(x.D), D \in B_Y.$ For the first
statement,  we show that the functional
\begin{eqnarray}  J(\mu) \equiv \int_{X\times Y} log \biggl( \frac {q(x,dy)}
{\int_X q(\xi,dy)\mu(d\xi)}\biggr)~ q(x,dy) \mu(dx) \label{eqEM43}\end{eqnarray} is concave. Define the measure
$\tilde \mu \in M(Y)$ by the convolution
$$\tilde \mu(D) \equiv \int_X q(x,D) \mu(dx), D\in B_Y. $$ Since, for each $D \in B_Y,$  $x \longrightarrow q(x,D)$ is
 continuous and bounded, with values from
$[0,1],$ this is well defined.   Choose any  $v \in M(Y)$ so that $q(x,\cdot) \prec v(\cdot)$ for all $x \in X.$
Such a choice is assured since $q(x,\cdot)$ is a regular Borel probability measure on $B_Y$ induced by  the
output process $\{y\}$ having  continuous sample paths.   Clearly $\tilde \mu \prec v$ also. Using this measure
we can express (\ref{eqEM43}) as the sum of two terms as follows,
\begin{eqnarray*}  J(\mu) &~& \equiv \int_{X\times Y} log \biggl( \frac {q(x,dy)} {\tilde
\mu(dy)}\biggr)~ q(x,dy) \mu(dx) \nonumber \\ &~& = \int_{X\times Y}\biggl\{  log \biggl(
\frac{q(x,dy)}{v(dy)}\biggr)  - log  \biggl( \frac {\tilde \mu(dy)}{v(dy)}\biggr)\biggr\}~  q(x,dy)
\mu(dx).\end{eqnarray*}  Then using Fubini's theorem  and interchanging the order of integration in the second
term we find that
\begin{eqnarray}  J(\mu) &=& \int_{X\times Y}\biggl\{ log \biggl( \frac{q(x,dy)}{v(dy)}\biggr)\biggr\}q(x,dy)
\mu(dx) - \int_Y log \biggl\{ \biggl( \frac {\tilde \mu(dy)}{v(dy)}\biggr)\biggr\}  \tilde \mu(dy)\nonumber \\
 &=& J_1(\mu) - J_2(\mu) \equiv J_1(\mu) - I_2(\tilde \mu).  \label{eqEM44}\end{eqnarray} Consider the function
 $$ \eta(x) \equiv \int_{Y} \log\biggl( \frac{q(x,dy)}{v(dy)}\biggr) q(x,dy).$$  We have already noted that for
 every $D \in B_Y$,
 $x \longrightarrow q(x,D)$ is continuous on  $X$ with values in $[0,1].$  This implies continuity of $\eta$.
 Indeed, $\eta$ is the uniform limit of the sequence of bounded continuous functions $\{\eta_m\}$ given by  $$ \eta_m(x) \equiv \sum_{i=1}^m \log\biggl(
 \frac{q(x,Y_i)}{v(Y_i)}\biggr) q(x,Y_i)$$ where  $\{Y_i\}$ is a partition of $Y$ by  pairwise disjoint
members $ Y_i \in B_Y.$
 Thus    $\mu
\longrightarrow J_1(\mu)$ is linear and  bounded. By definition, the second term  is  the relative entropy of
$\tilde \mu$ with respect to the measure $v.$  For an arbitrary but fixed $v\in M(Y)$, $\nu \longrightarrow
I_2(\nu) $ is  a strictly convex functional (\cite{dupuis97},Lemma 1.4.3, p36) on the set  $\{ \nu \in M(Y):
I_2(\nu) <\infty\}.$ This can be easily verified by use of Gibb's formula and strict convexity of the function
$\eta(\xi) \equiv \xi \log \xi, \xi \geq 0.$  Thus $I_2$ is strictly convex and hence $-I_2$ is strictly
concave.
  Combining these facts we conclude that
$\mu \longrightarrow J(\mu)$ is strictly concave proving the first part of the statement.  Now we consider the
question of  continuity. It is clear from the expression (\ref{eqEM44}) that $J_1$ is bounded linear and hence
continuous with respect to the weak topology. The functional $J_2,$ or equivalently $I_2,$ gives   the relative
entropy of $\tilde \mu$ with respect to the measure $v$. Again, it is well known that relative entropy is weakly
lower semicontinuous (\cite{dupuis97},Lemma 1.4.3, p36). Thus $J_2$ is weakly lower semicontinuous and hence
$-J_2$ is weakly upper semicontinuous. Hence the functional $J$ given by their sum  is weakly upper
semicontinuous proving the second part of the statement. Thus we conclude that $\mu \longrightarrow J(\mu) $ is
strictly concave and weakly upper semicontinuous. Next we verify that $\sup \{ J(\mu), \mu \in {\cal M}_{ad}\} <
\infty.$ Clearly it suffices to verify that $\sup \{ J(\mu), \mu \in S_r\} < \infty.$  An alternative expression
for the mutual information, well known in the literature,  ~(\cite{duncan}, Duncan,Theorem 2, p269),  and
(\cite{lipster78},Lipster-Shirayev, Theorem 16.3, p174) is given by $$ I({\cal X},{\cal Y}) = (1/2) E\biggl\{
\int_I \biggl( |F_t(x)|_{R^m}^2 - |\hat F_t(y)|_{R^m}^2\biggr) dt\biggr\} $$ where $\hat F_t(y) \equiv E\biggl\{
F_t(x)|{\cal F}_t^y\biggr\}$ with ${\cal F}_t^y$ denoting the smallest sigma algebra with respect to which the
process  $\{y(s), s \leq t\} $ is measurable.  This identity holds for all finite dimensional stochastic
differential equations like (\ref {eqEMS3}) and (\ref{eqEMS5}) with $F$ nonanticipative  and $F_t$ taking values
from $R^m.$ Under the assumption, $P \biggl\{ \int_I | F_t(x)|_{R^m}^2 dt < \infty \biggr\}=1, $ the proof is
identical. We give a brief outline. Let $\gamma(dx\times dy)$ with its marginals, $\mu(dx)$ and $\nu(dy),$  be
the measures as introduced  earlier and let $\beta(dy)$ denote the Wiener measure on $Y$. For the system
$$ dy = F_t(x) dt + dW, y(0) = 0,$$ it follows from absolute continuity  of $\gamma$ with respect to $\mu \times \beta,$
 and  $\nu$ with respect to $\beta,$ and  Girsanov measure substitution that  the Radon-Nikodym
derivative of $\gamma$ with respect to $\mu\times \beta$ is given by \begin{eqnarray*}  \bigl(\gamma(dx\times
dy)/\mu(dx)\times \beta(dy)\bigr) (x,y) &=& \exp\biggl\{ \int_I (F_t(x),dy) -(1/2) \int_I |F_t(x)|_{R^m}^2 dt\biggr\} \\
&=& \exp\biggl\{ (1/2) \int_I |F_t(x)|^2 dt + \int_I (F_t(x),dW) \biggr\}. \end{eqnarray*} Similarly, the RND of
$\nu$ with respect $\beta$ is given by \begin{eqnarray*}(\nu(dy)/\beta(dy))(y)=  \rho(y)  &\equiv& \exp\biggl\{
\int_I (\hat F_t(y),dy) -(1/2) \int_I |\hat F_t(y)|^2 dt \biggr\} \\ &=& \exp\biggl \{ \int_I (\hat
F_t(y),F_t(x))dt -(1/2) \int_I |\hat F_t(y)|^2 dt + \int_I <\hat F_t(y), dW)  \biggr\},
\end{eqnarray*} where $ \hat F_t(y) = E\bigl\{ F_t(x)|{\cal F}_t^y \bigr\}.$  Since by assumption $P \biggl \{
\int_I |F_t(x)|_{R^m}^2 dt < \infty \biggr\} =1$ it is clear that $\beta \{ \rho(y) = 0\} =0$ and hence $\beta\{
1/\rho(y) < \infty\} = 1.$ From these it follows that the RND of $\gamma$ with respect to the product measure
$\mu\times \nu$ is given by   \begin{eqnarray*} &~& (\gamma(dx\times dy)/\mu(dx)\times \nu(dy)\bigr)(x,y) =  \\
&~& ~~~~~~~~~~~~~~~= \exp\biggl\{ (1/2) \int_I |F_t(x)-\hat F_t(y)|_{R^m}^2 dt + \int(F_t(x)-\hat
F_t(y),dW)\biggr\}.\end{eqnarray*} Using this expression and the  definition of mutual information along with
standard properties of
 conditional expectations, we obtain
 \begin{eqnarray*} &~&  I({\cal X},{\cal Y}) \equiv E \biggl\{ \log \{ (\gamma(dx\times dy)/\mu(dx)\times \nu(dy)\bigr)(x,y)\}\biggr\}   =  \\
&~& ~~~~~~~~~~~~~~~= (1/2) E \biggl\{  \int_I |F_t(x)|^2 -\hat F_t(y)|^2 dt\biggr\} .\end{eqnarray*} This ends
the outline. Conditional expectation is a contraction map and so $I({\cal X},{\cal Y})\geq 0,$ as expected, and
we have $$ J(\mu) \equiv I({\cal X},{\cal Y}) \leq  (1/2) E \int_I |F_t(x)|_{R^m}^2 dt  \leq  (1/2) \int_X
\biggl\{ \int_I | F_t(x)|_{R^m}^2 dt\biggr\} \mu(dx).$$  Since $F$ is a bounded linear operator from $X$ to $Y$
there exists a finite positive number $K,$ independent of $t \in I,$  so that $ |F_t(x)|_{R^m}^2   \leq K ^2
|x|_{L_2([0,t],R^n)}^2 , t \in I.$  Hence it follows from the above inequality that for $\mu \in M_2(X)$,
$$ J(\mu) \leq (K^2T/2) \int_X |x|^2 \mu(dx) < \infty.$$ Thus on $S_r \subset M_2(X),$ we have
$$ \sup_{\mu \in S_r} J(\mu) \leq (K^2T^2 r/2) < \infty $$  proving that the functional $\mu
\longrightarrow J(\mu)$ is  bounded away from $+\infty$  on $S_r.$ Since $J$ is weakly upper semicontinuous and
bounded away from $+\infty$ and,  by our choice, the admissible set ${\cal M}_{ad} \subset S_r $ is weakly
compact, $J$ attains its supremum on ${\cal M}_{ad}.$  This proves the existence of a  $\mu^o \in {\cal M}_{ad}$
at which the capacity $C$ is attained, that is, $C = J(\mu^o).$ The uniqueness follows  from strict concavity
of  $ \mu\longrightarrow J(\mu)$ and convexity of ${\cal M}_{ad}.$  This completes the proof.  $\bullet$
\vskip6pt \noindent {\bf Remark 4.2 (Some Examples of Compact sets):} Here we present  some  simple  examples of
weakly compact subsets ${\cal M}_{ad}$ of the set $S_r.$ {\bf (E1):} Let $\{\mu_k\} \subset S_r $ be a family of
distinct measures, in the sense that $\mu_k \not \equiv \mu_m $ on $B_X$ for  $k \not = m,$ and $\Lambda $ a
subset of $\ell_1$ satisfying the following properties:
\begin{eqnarray*} &~& (1): \alpha_k \geq 0, \sum_{k=1}^{\infty} \alpha_k = 1  ~\hbox{for}~ \alpha \in \Lambda,
\\ &~& (2): \lim_{N \rightarrow \infty} \sum_{k \geq N} \alpha_k  = 0 , ~\hbox{uniformly in}~ \alpha \in
\Lambda.
\end{eqnarray*} Define the set $M_{\Lambda} \equiv \{ \mu \in M(X) : \mu = \sum_{k\geq 1} \alpha_k \mu_k, \alpha
\in \Lambda \}.$ The reader can easily verify that $M_{\Lambda} \subset S_r$ and it  is  uniformly tight and so
relatively weakly compact. Since $\Lambda$ is compact, $M_{\Lambda}$ is closed and hence it is    weakly
compact.
 {\bf (E2):} A variant of this example for which the  same conclusion
 holds is as follows. Let $X_0$ be  a   countable  dense subset of the closed ball $ B_{a}(X)$ of  $X$ of radius
$a \leq rT.$ Then the  set   $ M_{\Lambda}(X_0)  \equiv \{ \mu \in M(X) :  \mu = \sum_{k\geq 1} \alpha_k
 \delta_{x_k}, x_k \in X_0,
 \alpha \in \Lambda \}$ is a  weakly compact subset of $S_r.$ {\bf (E3):}  If $\Upsilon$ is any  uniformly tight subset
 of $S_r$, then the set ${\cal M}_{ad} \equiv  wc\ell(\Upsilon)$ given by the weak closure of $\Upsilon$   is  weakly
compact. {\bf (E4):} Let $D$ be a compact subset of the  Hilbert space $X$ and $M(D)= \{\mu \in  M(X):
\mu(X\setminus D) = 0\}.$ Clearly $M(D)$ is a weakly compact subset of $M(X).$  We may choose $D$ such  that $
S_r \cap M(D) \neq \emptyset $ and then take ${\cal M}_{ad} \equiv  S_r \cap  M(D).$  Since $X = L_2(I,R^n)$, it
follows from Sobolev embedding theorems that  for any finite interval $I$ and any $p \in [2,\infty)$ the
embedding $W^{1,p}(I,R^n) \hookrightarrow L_2(I,R^n)$ is continuous and compact. Thus a good example of a
compact set $D$ of $X$ is any closed bounded subset $D$ of $W^{1,p}(I,R^n).$
\section{Maximizing Source Measure}
In this section we wish to present necessary conditions that a (maximizing) measure, subject to energy
constraints and  determining the channel capacity, must satisfy. Such conditions are called necessary conditions
of optimality. In fact in the following result we have both necessary and sufficient conditions of optimality.
\vskip6pt \noindent {\bf Theorem 5.1} In order for $\mu^o \in {\cal M}_{ad} \subset M_2(X)$ to be the optimum
source measure, it is necessary and sufficient that the following inequality holds
\begin{eqnarray}&~&  \int_{X\times Y} \biggl\{ \log \biggl( \frac{N_G(\bar F(x),Q_1)(dy)}{\int_X
N_G(\bar F(\xi),Q_1)(dy) \mu^o(d\xi)} \biggr)\biggr\} N_G(\bar F(x),Q_1)(dy) \mu^o(dx)\nonumber \\ &~& ~~\geq
\int_{X\times Y} \biggl\{\log \biggl( \frac{N_G(\bar F(x),Q_1)(dy)}{\int_X N_G(\bar F(\xi),Q_1)(dy) \mu^o(d\xi)}
\biggr)\biggr\} N_G(\bar F(x),Q_1)(dy) \mu(dx)
 \label{eqEM45}\end{eqnarray} for all $\mu \in {\cal M}_{ad}.$\vskip6pt \noindent {\bf Proof}
  Let $\mu^o\in {\cal M}_{ad} $ denote the optimizer, that is,
$$  J(\mu^o) = C \equiv \sup \{ J(\mu), \mu \in  {\cal M}_{ad}\}.$$ Then for any $\mu \in  {\cal M}_{ad},$   define
$\mu^{\varepsilon} \equiv \mu^o + \varepsilon (\mu-\mu^o)$ for $\varepsilon \in (0,1).$  Since $ {\cal M}_{ad}$
is convex, it is clear that $\mu^{\varepsilon} \in {\cal M}_{ad}.$  Clearly  $\mu^o$ being  the maximal element,
we have $J(\mu^o) \geq J(\mu^{\varepsilon}).$ Let  $DJ(\mu)$ denote the Gateaux gradient of $J$ at $\mu$
whenever it exists. Then computing  the limit, $$ \lim_{\varepsilon \downarrow 0}\biggl\{ \frac{
J(\mu^{\varepsilon})-J(\mu^o)}{ \varepsilon} \biggr\} = <DJ(\mu^o),\mu-\mu^o>,$$ and noting that for optimality,
\begin{eqnarray}   <DJ(\mu^o),\mu-\mu^o> ~ \leq ~ 0, \label{eqEM46} \end{eqnarray}  it is easy to verify that
\begin{eqnarray} \int_{X\times Y} \biggl\{ 1 - \log\biggl( \frac{N_G(\bar F(x),Q_1)(dy)}{\int_X N_G(\bar
F(\xi),Q_1)(dy) \mu^o(d\xi)} \biggr) \biggr\} N_G(\bar F(x),Q_1)(dy) (\mu-\mu^o)(dx) \geq 0,\nonumber \\
\label{eqEM47}
\end{eqnarray} $ \forall~ \mu \in  {\cal M}_{ad}.$ The inequality (\ref{eqEM45}) now easily follows from (\ref{eqEM47}).
  For the sufficiency, recall that $\mu\rightarrow
J(\mu)$ is concave. Hence \begin{eqnarray}  <DJ(\mu),\nu-\mu> ~ \geq ~J(\nu)-J(\mu)~~ \forall \mu,\nu \in  {\cal
M}_{ad}. \label{eqEM48}\end{eqnarray} Taking $\mu = \mu^o$ it follows from this inequality  and (\ref{eqEM45}),
or equivalently  (\ref{eqEM46}),   that
\begin{eqnarray} 0 ~ \geq ~ <DJ(\mu^o),\nu-\mu^o> ~ \geq ~ J(\nu)-J(\mu^o)~~ \forall~~\nu \in
 {\cal M}_{ad} .\label{eqEM48}\end{eqnarray} and hence $J(\mu^o) \geq J(\nu) ~\forall~ \nu \in {\cal M}_{ad}.$
  This proves the
sufficiency of condition (\ref{eqEM45}) thereby  completing the proof. $\bullet $ \vskip6pt \noindent {\bf
Remark 5.2} ~  So far in sections 4 and 5,  we have assumed the source of electro-magnetic fields to be  noise
free. As seen in section 3.3, to include noisy source, we must  consider the  evolution equations (\ref{eqES1})
and (\ref{eqES3}) which are the  stochastic versions of Maxwell (field)  equations. With the  source noise
included, the results of Theorem 4.1 and Theorem 5.1 remain valid with the replacement of the Channel Kernel
$N_G(\bar F(x),Q_1)(\cdot)$ by  $N_G(\bar F(x),Q_1 + Q_2)(\cdot)$ where $Q_2$ is the covariance operator
associated with the source noise. For the distributed source model (\ref{eqES1}), the covariance operator $Q_2$
is given by
\begin{eqnarray} (Q_2\xi,\xi) = \int_{I\times I} <K_2(t,\tau) \xi(t),\xi(\tau)> dt d\tau
\end{eqnarray} where the kernel $K_2$ is given in terms of the system parameters as follows:
 \begin{eqnarray} K_2(t,\tau) =  \int_0^{t\wedge \tau} dr R(t,r)
R^*(\tau,r), \end{eqnarray} with $R(t,r), 0 \leq r \leq t \leq T,$ given by  \begin{eqnarray}  R(t,r) \equiv
{\cal G} \Gamma\int_r^t {\cal S}(\theta-r) \sigma d\theta = \int_r^t {\cal G} \Gamma {\cal S}(\theta-r) \sigma
d\theta.\end{eqnarray} The last identity follows from the fact that ${\cal G}\Gamma$ is a bounded linear
operator  from the state space ${\cal H}$ to $R^m$  and hence closed and so commutes with the integral
operation.
 Similar expressions can be derived for the boundary source.  These  conclusions  are  based on the properties of
 stochastic convolutions,
see (\cite{daprato92},Theorem 5.2, p119) 
which use stochastic Fubini's theorem~(\cite{daprato92},Theorem 4.18, p109).

\vskip6pt \noindent {\bf  Remark  5.3 (Numerical Algorithm for Computation):} Based on the necessary (and
sufficient) conditions of optimality as presented above, we can develop a gradient based algorithm for numerical
computation. In particular, for simple sources like those of examples, (E1) and (E2), the functional $J(\mu)$ on
${\cal M}_{ad}$ can be redefined as being a functional on  $\Lambda$ and, with slight abuse of notation, we may
denote it by $J(\alpha).$ Our problem is to find an $\alpha^o \in \Lambda$ at which $J$ attains its supremum.
For the source (E1), it follows from the necessary conditions that in order for $\alpha^o \in \Lambda $ to be
the optimal distribution of weights assigned to the family of measures $\{\mu^k\}$, it is necessary and
sufficient that we have,
\begin{eqnarray} \sum_{k\geq 1} (\alpha_k -\alpha_k^o) \int_X   L_{\alpha^o}(x)~  \mu^k(dx) \geq 0 , ~~\forall~~
\alpha \in \Lambda, \end{eqnarray} where $$ L_{\alpha^o} (x) \equiv \int_Y \log\biggl ( q(x,dy)/ \sum_{k \geq 1}
\alpha_k^o ~ \nu^k(dy)\biggr) q(x,dy),~ \nu^k(dy) = \int_X q(x,dy) \mu^k(dx). $$  For the  source (E2), this
further reduces to $$ \sum_{k\geq 1} (\alpha_k-\alpha_k^o) L_{\alpha^o}(x_k) \geq 0, ~\forall~ \alpha \in
\Lambda,$$ where $$L_{\alpha^o}(x_k) \equiv \int_Y \log\biggl( q(x_k,dy)/\sum_{k\geq 1}\alpha_k^o~
q(x_k,dy)\biggr) q(x_k,dy).$$  For these simple sources, one can write a gradient based numerical algorithm for
finding the optimum source. Considering  the first source, the gradient of $J$ at the n-th iteration is given by
$$ \triangle J(\alpha^n) =\{ \triangle_kJ(\alpha^n) =  \int L_{\alpha^n}(x) \mu^k(dx), k = 1,2,\cdots \}.$$
Choose  $$  \alpha_k^{n+1} \equiv \alpha_k^n + \varepsilon_n \triangle_k J(\alpha^n), k \in N, $$ with
$\varepsilon_n
>0$ sufficiently small so that $\alpha^{n+1} \in \Lambda.$   Using this $\alpha^{n+1}$, we compute the objective functional
 giving  $$ J(\alpha^{n+1}) =  J(\alpha^n) + \varepsilon \sum_{k=1}^{\infty} (\triangle_k
J(\alpha^n))^2 + 0(\varepsilon).$$ For $\varepsilon_n >0$ sufficiently small the series converges guaranteeing
improvement of  the objective functional at each step. Similar conclusion holds for the source (E2) with the
gradient vector given by $$ \triangle J(\alpha^n) =\{ \triangle_kJ(\alpha^n) =  L_{\alpha^n}(x_k) , k =
1,2,\cdots \}.$$.

\section{Conclusion} We have presented a complete dynamic model for MIMO channels
(wave guides, cavities)  based on deterministic as well as stochastic Maxwell's equations. To the best  of our
knowledge, this  formulation  has not been considered in the literature. Both distributed and boundary sources
have been considered. Proof of existence of maximizing source measure subject to power constraints has been
presented. Optimality conditions have been developed which can be used for numerical computations as indicated
in remark 5.3 for simple source spaces. The model used for sensors  in the above formulation can be extended to
cover linear (sensor) dynamics without any difficulty. This requires inclusion of the  associated transition
operator for the construction of the output map $F_t(x).$
 It would be interesting to study similar problems with nonlinear sensor dynamics.
\par\noindent {\bf Acknowledgement} (1):  The first author would like to thank his colleague,  Professor D. A. McNamara, for many valuable
discussions on physical properties of waveguide channels. (2):  Also thanks are due to National Science and
Engineering Research Council of Canada for partially supporting this research under grant no A7109.




\end{document}